\documentclass[iop]{emulateapj}

\usepackage{url}
\usepackage{hyperref}

\usepackage{graphicx}        
\usepackage{bm}              
\usepackage{natbib}
\usepackage{color}
\usepackage{amsmath}

\usepackage{hhline}

\usepackage[mediumspace,Gray,squaren]{SIunits}

\shorttitle{ACT: Beams and Planet Temperatures}
\shortauthors{Hasselfield, Moodley, et al.}

\newcommand{\arone}{148\,GHz}
\newcommand{\artwo}{218\,GHz}
\newcommand{\arthree}{277\,GHz}

\newcommand{\artwoX}{218\,GHz}

\newcommand{\arlist}{148, 218, and 277\,GHz}

\newcommand{\commentx}[1]{}

\renewcommand{\vec}[1]{\mbox{\boldmath$#1$}} 
\newcommand{\unitvec}[1]{\ensuremath{\hat{\vec{#1}}}}

\newcommand{\ra}[3]   
   {\makebox[1.5em][r]{#1}\makebox[1.5em][r]{#2} \makebox[2em][r]{#3}}

\newcommand{\hms}[3]  
   {${#1}^{\mathrm{h}}{#2}^{\mathrm{m}}{#3}^{\mathrm{s}}$}

\newcommand{\hmin}[2]  
   {\ensuremath{{#1}^{\mathrm{h}}{#2}^{\mathrm{m}}}}
   
\newcommand{\hours}[1]  
   {\ensuremath{{#1}^{\mathrm{h}}}}

\newcommand{\dms}[3]  
   {\ensuremath{{#1}\degree{#2}\arcminute{#3}\arcsecond}}

\newcommand{\dm}[2]  
   {\ensuremath{{#1}\degree{#2}\arcminute}}

\newcommand{\ukcmb}  
           {\ensuremath{\micro \kelvin_\mathrm{cmb}}}

\newcommand{\uk}  
           {\ensuremath{\micro \kelvin}}

\newcommand{\fdeg} 
           {\hbox{$.\!\!^{\circ}$}}

\usepackage{color}

\newcommand{\omitt}[1]{ }

\newcommand{\hide}[1]{\editorial{\emph{Hidden text...}} }

\definecolor{orange}{rgb}{1,0.3,0}
\definecolor{purple}{rgb}{1,0,1}
\newcommand{\editorial}[1]{\textcolor{orange}{#1} }

\hypersetup{ colorlinks,
   linkcolor=blue,
   filecolor=blue,
   urlcolor=blue,
   citecolor=blue}

\newcommand\mytitle{Oh title, my title.}
\newcommand\mycaption{Oh caption, my caption.}

\begin{document}

\title{The Atacama Cosmology Telescope: Beam Measurements and the
  Microwave Brightness Temperatures of Uranus and Saturn}


\author{
Matthew~Hasselfield\altaffilmark{1,2},
Kavilan~Moodley\altaffilmark{3},
J.~Richard~Bond\altaffilmark{4},
Sudeep~Das\altaffilmark{5,6},
Mark~J.~Devlin\altaffilmark{7},
Joanna~Dunkley\altaffilmark{8},
Rolando~D\"{u}nner\altaffilmark{9},
Joseph~W.~Fowler\altaffilmark{10,11},
Patricio~Gallardo\altaffilmark{9,11},
Megan~B.~Gralla\altaffilmark{12},
Amir~Hajian\altaffilmark{4},
Mark~Halpern\altaffilmark{2},
Adam~D.~Hincks\altaffilmark{4},
Tobias~A.~Marriage\altaffilmark{12},
Danica~Marsden\altaffilmark{13,7},
Michael~D.~Niemack\altaffilmark{10,11,14},
Michael~R.~Nolta\altaffilmark{4},
Lyman~A.~Page\altaffilmark{11},
Bruce~Partridge\altaffilmark{15},
Benjamin~L.~Schmitt\altaffilmark{7},
Neelima~Sehgal\altaffilmark{16},
Jon~Sievers\altaffilmark{1,4},
Suzanne~T.~Staggs\altaffilmark{11},
Daniel~S.~Swetz\altaffilmark{10,7},
Eric~R.~Switzer\altaffilmark{4},
Edward~J.~Wollack\altaffilmark{17}
}
\altaffiltext{1}{Department of Astrophysical Sciences, Peyton Hall, 
Princeton University, Princeton, NJ 08544, USA}
\altaffiltext{2}{Department of Physics and Astronomy, University of
British Columbia, Vancouver, BC, V6T 1Z4, Canada}
\altaffiltext{3}{Astrophysics and Cosmology Research Unit, School of
Mathematics, Statistics \& Computer Science, University of KwaZulu-Natal,
Durban, 4041, South Africa}
\altaffiltext{4}{Canadian Institute for Theoretical Astrophysics, University of
Toronto, Toronto, ON, M5S 3H8, Canada}
\altaffiltext{5}{High Energy Physics Division, Argonne National Laboratory,
9700 S Cass Avenue, Lemont, IL 60439, USA}
\altaffiltext{6}{Berkeley Center for Cosmological Physics, LBL and
Department of Physics, University of California, Berkeley, CA 94720, USA}
\altaffiltext{7}{Department of Physics and Astronomy, University of
Pennsylvania, 209 South 33rd Street, Philadelphia, PA 19104, USA}
\altaffiltext{8}{Department of Astrophysics, Oxford University, Oxford, OX1
3RH, UK}
\altaffiltext{9}{Departamento de Astronom{\'{i}}a y Astrof{\'{i}}sica, 
Facultad de F{\'{i}}sica, Pontific\'{i}a Universidad Cat\'{o}lica,
Casilla 306, Santiago 22, Chile}
\altaffiltext{10}{NIST Quantum Devices Group, 325
Broadway Mailcode 817.03, Boulder, CO 80305, USA}
\altaffiltext{11}{Joseph Henry Laboratories of Physics, Jadwin Hall,
Princeton University, Princeton, NJ 08544, USA}
\altaffiltext{12}{Dept. of Physics and Astronomy, The Johns Hopkins University, 3400 N. Charles St., Baltimore, MD 21218-2686, USA}
\altaffiltext{13}{Department of Physics, University of California Santa Barbara,
CA 93106, USA}
\altaffiltext{14}{Department of Physics, Cornell University, Ithaca, NY 14853, USA}
\altaffiltext{15}{Department of Physics and Astronomy, Haverford College,
Haverford, PA 19041, USA}
\altaffiltext{16}{Department of Physics and Astronomy, Stony Brook, NY 11794-3800, USA}
\altaffiltext{17}{NASA/Goddard Space Flight Center,
Greenbelt, MD 20771, USA}


\newcommand\TUoneRJ{103.2~K}
\newcommand\TUoneB{106.7~K}
\newcommand\TUoneE{2.1\%}
\newcommand\TUoneBE{$106.7 \pm 2.2$~K}

\newcommand\TUtwoRJ{95.0~K}
\newcommand\TUtwoB{100.1~K}
\newcommand\TUtwoE{3.2\%}
\newcommand\TUtwoBE{$100.1 \pm 3.1$~K}

\newcommand\TSoneBE{$137.3 \pm 3.2$~K}
\newcommand\TStwoBE{$137.3 \pm 4.7$~K}

\begin{abstract}
We describe the measurement of the beam profiles and window functions for
the Atacama Cosmology Telescope (ACT), which operated from 2007 to 2010
with kilo-pixel bolometer arrays centered at \arlist{}.  Maps of Saturn are
used to measure the beam shape in each array and for each season of
observations.  Radial profiles are transformed to Fourier space in a way
that preserves the spatial correlations in the beam uncertainty, to derive
window functions relevant for angular power spectrum analysis.  Several
corrections are applied to the resulting beam transforms, including an
empirical correction measured from the final CMB survey maps to account for
the effects of mild pointing variation and alignment errors.  Observations
of Uranus made regularly throughout each observing season are used to
measure the effects of atmospheric opacity and to monitor deviations in
telescope focus over the season.  Using the WMAP-based calibration of the
ACT maps to the CMB blackbody, we obtain precise measurements of the
brightness temperatures of the Uranus and Saturn disks at effective
frequencies of 149 and 219 GHz.  For Uranus we obtain thermodynamic
brightness temperatures $T_{\rm U}^{149} = $ \TUoneBE{} and $T_{\rm
  U}^{219} = $ \TUtwoBE{}. For Saturn, we model the effects of the ring
opacity and emission using a simple model and obtain resulting (unobscured)
disk temperatures of $T_{\rm S}^{149} = $ \TSoneBE{} and $T_{\rm S}^{219} =
$ \TStwoBE{}.

\end{abstract}



\newcommand{\secref}[1]{Section~\ref{#1}}

\section{Introduction}
\setcounter{footnote}{0} 

The Atacama Cosmology Telescope (ACT) is a 6 m, off-axis telescope designed
for millimeter wavelength observations of the cosmic microwave background
(CMB) with arcminute resolution \citep{fowler/etal:2007}.  The telescope is
located in Northern Chile, at an elevation of 5190~m, where atmospheric
conditions are excellent for microwave observations from April through
December. From 2008 to 2010, ACT observed the sky with bolometer arrays
operating in frequency bands centered at \arone{}, \artwo{} and \arthree{}.
The choice of bands provides the ability, in principle, to spectrally
discriminate between signal from the CMB, the Sunyaev-Zel'dovich effect
\citep[SZE; ][]{sunyaev/zeldovich:1970}, and dusty sources.  The \arone{}
array was also in operation during a short observing season at the end of
2007.

The ACT scientific program has produced measurements of the CMB
angular power spectrum at multipoles spanning $\ell \approx 300$ to 10000
\citep{das/etal:2013, dunkley/etal:2013}, probing the primary and secondary
anisotropies \citep{sievers/etal:2013}.  ACT has produced source catalogs
\citep{marriage/etal:2011b}, and galaxy cluster samples reaching to high
redshifts \citep{marriage/etal:2011, hasselfield/etal:prep}.

The telescope beam acts as a low-pass spatial filter on the astrophysical
signal from the sky.  The effects of this filtering are relevant, to
varying degrees, to all ACT science results.  Interpretation of the angular
power spectrum, in particular, relies on its decomposition into components
that are correlated across large ranges in angular scale.  The extraction
of cosmological parameters associated with such components depends on an
understanding of the beam shape and the covariant features of the window
function uncertainty.

In this work we describe the processing of planet observations to obtain
the telescope beams, and thus the instrument window function, for each
detector array for each season of operation.  We also discuss the use of
Uranus observations to monitor beam shape and calibration.  Because the ACT
maps are ultimately calibrated to the CMB blackbody via the dipole
calibration of WMAP \citep{jarosik/etal:2011}, we obtain precise
measurements of the Uranus and Saturn disk temperatures without resorting
to interpolation across unknown source spectra or other calibration
transfer standards.

A detailed description of the instrument can be found in
\cite{swetz/etal:2011}, and the main data reduction and map-making pipeline
is described in \cite{dunner/etal:2013}.  \cite{hincks/etal:2010} present
an analysis of the ACT beams from the 2008 season; the present approach
extends this earlier analysis.

In \secref{sec:observations} of this paper we discuss the planet
observations and maps.  In \secref{sec:beams} we obtain the telescope
window function through an analysis of the planet maps and the application
of corrections for several systematic effects.  In
\secref{sec:calibration}, we interpret the apparent brightness of Uranus to
calibrate the sensitivity of the telescope to atmospheric water vapor and
focus variation.  In \secref{sec:brightness} we use the CMB-based
calibration of the ACT maps to obtain absolutely calibrated measurements of
Uranus and Saturn brightness temperatures.
\\

\section{Observations and Mapping}
\label{sec:observations}

\subsection{Observations}

The ACT observation strategy is to scan over small ranges of azimuth angle
while keeping the boresight altitude fixed at $\approx 50^\circ$ to
minimize systematic effects due to altitude variation.  The CMB fields are
observed as they rise and set through two central azimuth pointings.
Planet observations are made at this same telescope altitude angle, by
briefly interrupting CMB survey scans and re-pointing in azimuth to the
planet location.

ACT observed a planet approximately once per night of operation in
2007--2010, but only high quality observations of Saturn and Uranus are
considered in this analysis.  Saturn and Uranus were preferentially
targeted, as Saturn's brightness makes it useful for beam profile
measurements, and Uranus' stable brightness is a convenient calibration
source.  While Uranus was visible throughout each observing season, Saturn
was available only at the beginning and end of each season.  Jupiter and
Mars were, respectively, too bright and too low in the sky to be useful for
calibration work.  In Table~\ref{tab:observations} we list the number of
high quality Uranus and Saturn observations achieved in each season.

\renewcommand\mytitle{
Number of Saturn and Uranus observations selected for
calibration purposes for each season and array.
}

\renewcommand\mycaption{
While Uranus was visible throughout each observing season, Saturn was
available only near the beginning and end of each season.  The number of
successful planet observations is lower at higher frequencies because of
increased sensitivity in these arrays to atmospheric contamination.
} \begin{deluxetable}{l c c c c}
\tablecaption{\mytitle{}}
\tablehead{
    & 2007 & 2008 & 2009 & 2010
}
\startdata
Saturn & & & & \\
 \quad \arone{}& \phantom{0}22& \phantom{0}22& \phantom{0}13& \phantom{0}35\\
 \quad \artwo{}& --& \phantom{0}23& \phantom{00}9& \phantom{0}28\\
 \quad \arthree{}& --& \phantom{0}16& \phantom{00}5& \phantom{0}23\\
\cline{1-5} \vspace{-.1cm} & & \\
Uranus & & & & \\
 \quad \arone{}& \phantom{0}16& \phantom{0}37& \phantom{0}94& 113\\
 \quad \artwo{}& --& \phantom{0}33& \phantom{0}75& 102\\
 \quad \arthree{}& --& \phantom{0}21& \phantom{0}16& \phantom{0}74
\enddata
\label{tab:observations}
\tablecomments{\mycaption{}}
\end{deluxetable}

\subsection{Planet Maps}

As described in \cite{dunner/etal:2013}, maps are made from the ACT time
stream data according to a maximum likelihood technique in which the noise
is described in the time domain while the signal is assumed to be spatially
coherent \citep{tegmark:1997a}.  The bolometer time stream data for planet
observations are pre-processed in a similar fashion to the CMB survey data.
The time stream data are calibrated, based on detector load curves, after
deconvolving the effects of the detector time constants and low-pass
readout filtering.  Detectors are then screened for quality based on their
projections onto a common-mode computed using a fiducial set of
well-behaved detectors.  Detector screening is performed with the planetary
signal masked out.  Maps are made using a dedicated code that is optimized
for high resolution mapping of single observations; we have confirmed that
the main season map pipeline code produces compatible results when run on
the same data.  Each map is solved iteratively, converging after fewer than
10 iterations.  The noise model is designed to remove common-mode signal at
frequencies below 0.2 Hz, corresponding to angular scales that are much
larger than the array size.  Because high-pass filtering of this kind may
suppress signal at large angular scales, we study the transfer function of
the map making solutions in \secref{sec:sys_mapping}.

\subsection{Map Selection}

While planets were observed almost every night, observations do not always
yield successful maps.  The primary reason for failed mapping is a low
number of live detectors during periods of high sky loading.  This is a
greater problem at higher frequencies, and for the \arthree{} array in
particular leads to many maps with incomplete sampling of the planet
signal; such maps are discarded.

Planet observations are sometimes made after sunrise.  While such
observations allow us to study the magnitude of pointing and beam focus
changes due to thermal deformation of the telescope structure, they are
excluded from the present analysis.  The sunrise cut is also applied to CMB
observations.

\begin{figure*}[ht]
\begin{center}
\resizebox{\textwidth}{!}{\plotone{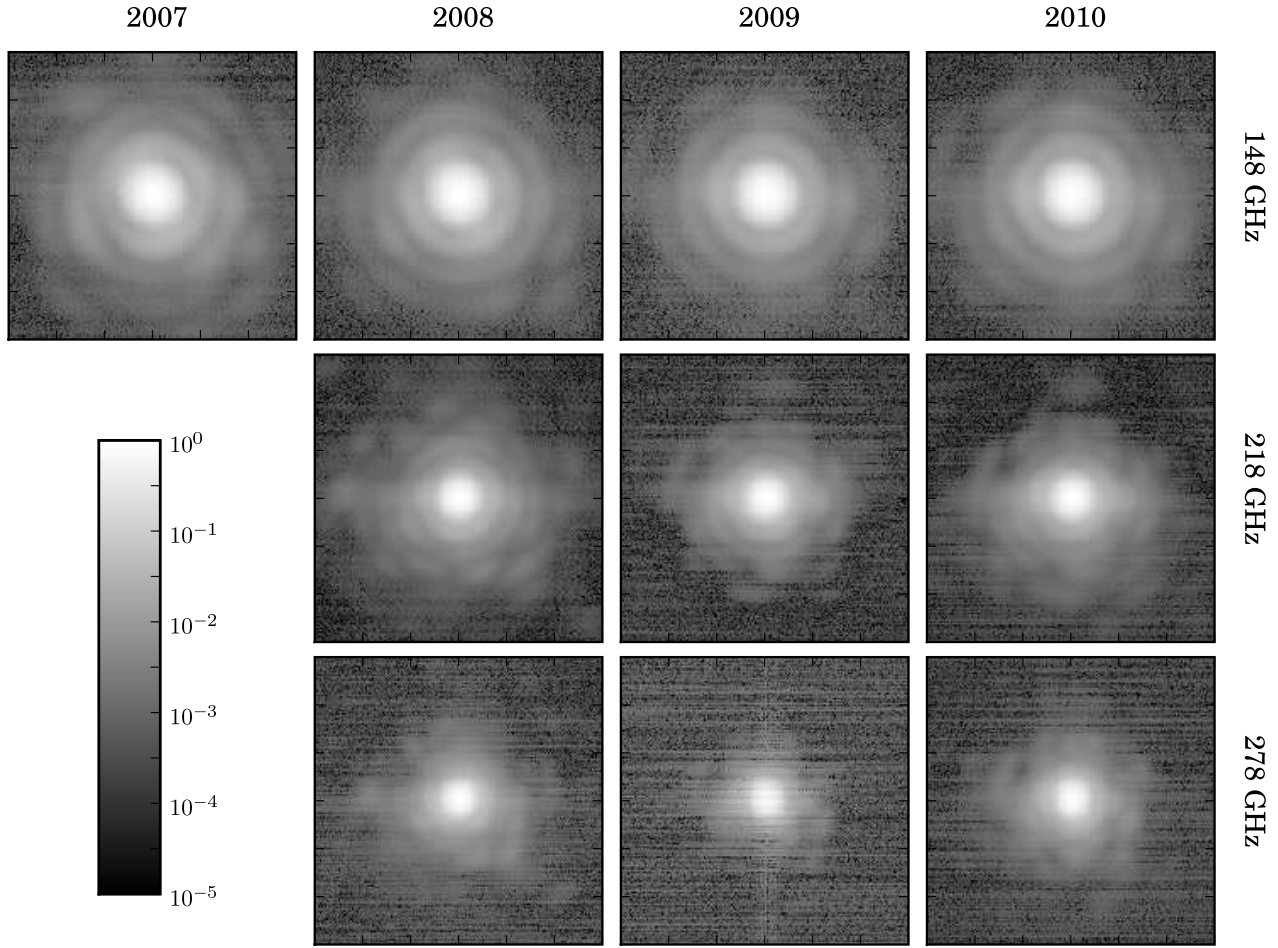}}
\caption{ Beam maps for each season and array, formed from the average of
  selected Saturn maps.  Gray scale is logarithmic.  Each panel is
  12\arcmin{} by 12\arcmin{}.Each contributing Saturn map incorporates all
  live detectors in the array.  The well-resolved Airy rings indicates that
  the relative detector offsets are accurately known.  The frequency labels
  correspond to the effective frequencies for CMB spectrum radiation
  \citep{swetz/etal:2011}.}
\label{fig:beam_maps}
\end{center}
\end{figure*}

A small number of maps are cut due to having substantially higher noise
levels than other maps of the same planet for the same array and season.
The remaining maps are used to establish the telescope beam and calibration
parameters, as described in the following sections.  Season mean beam maps,
obtained by aligning and averaging the selected Saturn maps for each season
and array, are shown in Figure~\ref{fig:beam_maps}.  The Airy pattern is
easily seen in the maps at \arone{} and \artwo{}.  The horizontal streaks
are parallel to the telescope scan direction.
\\

\section{Beams}
\label{sec:beams}

\renewcommand\unitvec[1]{{\mathbf{#1}}}
\newcommand\ellv{{\vec{\ell}}}
\newcommand\ellmax{\ell_{\rm max}}
\newcommand\Tt{\widetilde{T}}

In this section we obtain measurements of the telescope beams for each
season and array.  Saturn is bright and much smaller than the beam solid
angle, permitting the characterization of our beams with high
signal-to-noise ratio.  Saturn maps are reduced to radial profiles, which
are then modeled with appropriately chosen basis functions and transformed
to Fourier space.  The resulting transform, the beam modulation transfer
function, is corrected for a number of systematic effects to produce a
window function and $\ell$-space covariance suitable for use with the ACT
CMB survey maps.

\subsection{Radial Profiles}
\label{sec:radial_profiles}

While the instantaneous ACT beams are slightly ($< 10\%$) elliptical in
cross-section, we analyze them as though they were circular, and work
exclusively with the ``symmetrized'' (i.e., azimuthally averaged) radial
profiles.  The resulting beam measurements are suitable for any analysis
that incorporates a similar simple azimuthal average of the signal under
study.  In particular, the cross-linking of CMB survey observations is such
that the telescope beam contributes to the map, with roughly equal weight,
at two orientations that differ in rotation by approximately 90 degrees;
the net effective beam is very nearly circular.  The goal of our analysis
is thus to characterize the symmetrized beam.

Radial profiles are obtained for each Saturn observation by averaging map
data in radial annuli of width 9\arcsec{}, producing average profile
measurements $y_i$ at radii $\theta_i$.  On scales smaller than a few
arcminutes, the covariant noise is sub-dominant to the planetary signal and
thus the shape of the core beam is very well measured by even a single
observation of Saturn.  However, at larger angular scales there is
non-negligible contamination from the atmosphere.  This means that the
asymptotic behavior of the beam cannot be separated from the fluctuating
background without making some assumptions about the beam behavior far from
the main lobe.

The illumination of the ACT optics is controlled by a cryogenic Lyot stop
at an image of the circular exit pupil; this leads to a beam shape
described by an Airy pattern (for monochromatic radiation) and to the
expectation that the beam pattern (including the effects of finite detector
size and spectral response) will decay asymptotically as $1/\theta^3$,
where $\theta$ is the angle from the beam peak.  We fit this model to the
radial profile data in order to fix the background level of the map and to
provide a model for extrapolating the beam to large angles.

\begin{figure}[ht*]
\begin{center}
\resizebox{3.5in}{!}{\plotone{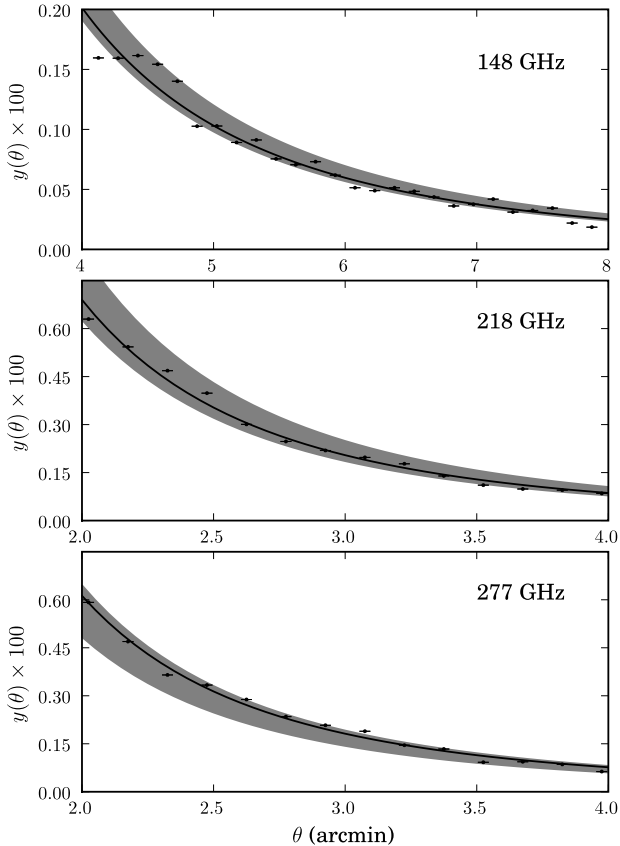}}
\caption{ Wing model fitted to binned radial profile data.  Points shown,
  and the best-fit model (solid line) are from a typical individual Saturn
  observation for each array.  A baseline is fit simultaneously and has
  been removed from the data points.  The shaded area represents the
  standard deviation of all individual best-fits to wing models for the
  2010 season. }
\label{fig:wing_fits}
\end{center}
\end{figure}

The background level of the map and the large-angle behavior of the beam
are measured by fitting the model $y(\theta) = A \times
(1\arcmin{}/\theta)^3 + c$ to the binned profile data in the range
$\theta_A < \theta_i < \theta_B$, with the range chosen such that the
profile measurements have fallen to below 1\% of the peak but are still
signal dominated.  The points $y_i$ with $\theta_i < \theta_A$, corrected
for the background level $c$ and renormalized so that $y(\theta=0) \approx
1$, constitute the ``core'' of the beam profile.  The radial profile beyond
$\theta_A$ is referred to as the ``wing.''  The fit ranges used for each
array, and the amplitude of the beam at $\theta_A$ relative to the peak,
are given in Table~\ref{tab:beam_props}.  An example of the wing fit for
each array is shown in Figure~\ref{fig:wing_fits}.

\renewcommand{\mytitle}{ Summary of beam parameters by array and season.  }
\renewcommand{\mycaption}{ Radial profiles are obtained within a distance
  $\theta_A$ from the peak; wing and baseline parameters are fit to data
  over the range $\theta_A$ to $\theta_B$.  The level $B(\theta_A)$ of the
  beam, relative to the peak, at the edge of the wing fit region is shown
  for reference.  Parameters $\ellmax{}$ and $n_{\rm mode}$ describe the
  basis functions used for fitting a beam model to the radial profile data
  in the beam core.  The solid angle and FWHM of the symmetrized
  instantaneous telescope beam are provided, along with the ellipticity
  (defined as the ratio of major to minor axes of the unsymmetrized mean
  beam).  The solid angles for the 2008 season may be compared to the
  values from the independent analysis of \citet{hincks/etal:2010}.  }
\begin{deluxetable*}{c c c c c c c c c c c c c c c}
\tablecaption{\mytitle{}}
\tablehead{
 &  & \multicolumn{3}{c}{Wing Fit (\S \ref{sec:radial_profiles})} &  & \multicolumn{2}{c}{Transform Fit (\S \ref{sec:harmonic_transforms})} &  & \multicolumn{3}{c}{Beam Properties} &  & \multicolumn{1}{c}{Hincks et al}\\
 \cline{3-5} \cline{7-8} \cline{10-12} \cline{14-15} \vspace{-.2cm}  & & & & & & & & \\
 Array & Season  & $\theta_A$ & $\theta_B$ & $B(\theta_A)$  &  & $\ell_{\rm max}$ & $n_{\rm mode}$  &  & Solid angle & FWHM & Ellip. & & Solid angle\\
 &  &  &  &  $(10^{-3})$ &  &  &  &  & ($10^{-9}$ sr)  & (arcmin) & &  & ($10^{-9}$ sr)
}
\startdata
\arone{} & 2007 & 4\arcmin{} & 8\arcmin{}  & 2.40 $\pm$ 0.05  &  & 21900 & 11 &  & $224.5 \pm 1.4$ & 1.364\arcmin{} & 1.098  & & -- \\
 & 2008 & 4\arcmin{} & 8\arcmin{}  & 2.03 $\pm$ 0.04  &  & 18400 & 9 &  & $216.6 \pm 1.6$ & 1.373\arcmin{} & 1.065  & & $218.2 \pm 4.0$ \\
 & 2009 & 4\arcmin{} & 8\arcmin{}  & 2.15 $\pm$ 0.09  &  & 16400 & 8 &  & $215.8 \pm 2.6$ & 1.378\arcmin{} & 1.039  & & -- \\
 & 2010 & 4\arcmin{} & 8\arcmin{}  & 2.13 $\pm$ 0.04  &  & 15800 & 9 &  & $215.4 \pm 1.8$ & 1.381\arcmin{} & 1.033  & & -- \\
\cline{1-14} \vspace{-.1cm} & & \\
\artwo{} & 2008 & 2\arcmin{} & 4\arcmin{}  & 6.54 $\pm$ 0.12  &  & 25500 & 6 &  & $116.5 \pm 1.1$ & 1.092\arcmin{} & 1.021  & & $118.2 \pm 3.0$ \\
 & 2009 & 2\arcmin{} & 4\arcmin{}  & 7.07 $\pm$ 0.29  &  & 25800 & 6 &  & $123.9 \pm 1.7$ & 1.057\arcmin{} & 1.026  & & -- \\
 & 2010 & 2\arcmin{} & 4\arcmin{}  & 7.34 $\pm$ 0.21  &  & 25100 & 6 &  & $125.0 \pm 1.8$ & 1.015\arcmin{} & 1.027  & & -- \\
\cline{1-14} \vspace{-.1cm} & & \\
\arthree{} & 2008 & 2\arcmin{} & 4\arcmin{}  & 5.49 $\pm$ 0.12  &  & 28400 & 7 &  & $\phantom{0}98.1 \pm 1.9$ & 0.869\arcmin{} & 1.049  & & $104.2 \pm 6.0$ \\
 & 2009 & 2\arcmin{} & 4\arcmin{}  & 5.22 $\pm$ 0.26  &  & 32100 & 7 &  & $\phantom{0}95.2 \pm 4.0$ & 0.879\arcmin{} & 1.118  & & -- \\
 & 2010 & 2\arcmin{} & 4\arcmin{}  & 5.65 $\pm$ 0.18  &  & 28600 & 7 &  & $\phantom{0}98.9 \pm 2.0$ & 0.870\arcmin{} & 1.103  & & -- 
\enddata
\label{tab:beam_props}
\tablecomments{\mycaption{}}
\end{deluxetable*}

For each array, the multiple observations of Saturn in each season are
combined by taking the mean of their core profile points and wings.  The
season mean beam profile is thus described by points $\bar{y}_i$ (for each
$\theta_i < \theta_A$) and a mean wing fit parameter $\bar{A}$.  The full
covariance matrix of $\{\bar{y}_1,...,\bar{y}_n,\bar{A}\}$ is computed from
the ensemble of Saturn profiles and, as described in the next section, is
used to propagate covariant error on large angular scales into the
$\ell$-space beam covariance matrix.

The solid angle and FWHM of the beam (as well as the approximate
ellipticity of the non-symmetrized telescope beam) are shown in
Table~\ref{tab:beam_props}.  Solid angles may be compared to the values
presented in \cite{hincks/etal:2010}.  The reduction in solid angle
uncertainty is primarily due to our choice to fit the wing over a range of
radii closer to the beam center.  We have doubled the error from its formal
value to account for systematic variation in solid angle results as
different fitting ranges are used.

\subsection{Harmonic Transforms and Window Functions}
\label{sec:harmonic_transforms}

Sky temperature data may be compared to predictions based on cosmological
models through statistical measures such as the angular power spectrum of
temperature fluctuations.  The effect of the observing strategy and
telescope beam on the measurement of correlation functions in CMB data has
been discussed in, e.g., \cite{white/srednicki:1995}.  For ACT, it is
important to understand the beam and its covariance over a broad range of
angular scales.

A Gaussian random temperature field $T(\unitvec{n})$ at position
$\unitvec{n}$ on the sphere may be characterized by coefficients $C_\ell$
of the decomposition of its auto-correlation function:
\begin{align}
  \Big\langle T(\unitvec{n}) T(\unitvec{n'}) \Big\rangle & =
  \sum_{\ell=1}^\infty \frac{2\ell+1}{4\pi}
  C_\ell P_\ell(\unitvec{n}\cdot \unitvec{n'}),
\end{align}
where $P_\ell$ are Legendre polynomials.  Measurements
$\widetilde{T}(\unitvec{n})$ of the temperature field with a given
telescope and observing strategy will differ from the true sky temperature;
for ACT the primary effects are an uncertain calibration (which we neglect
for now) and the telescope beam.  For the case of an azimuthally symmetric
beam that does not vary in time or with position on the sky, the
auto-correlation function of the observed temperature map
$\Tt{}(\unitvec{n})$ will be related to the $C_\ell$ describing the
$T(\unitvec{n})$ by \citep[e.g.,][]{white/srednicki:1995}
\begin{align}
  \left\langle \Tt{}(\unitvec{n}) \Tt{}(\unitvec{n'}) \right\rangle & =
  \sum_\ell \frac{2\ell+1}{4\pi}
  C_\ell P_\ell(\unitvec{n}\cdot \unitvec{n'}) B_\ell^2,
\end{align}
where $B_\ell$ are the multipole coefficients of the beam, defined by
\begin{align}
  B(\theta) & = \sum_{\ell=1}^\infty B_\ell \sqrt{\frac{(2\ell + 1)}{4\pi}} 
  P_\ell(\cos\theta).
\end{align}
In this context, $B_\ell^2$ is referred to as the window function.

The ACT beams are sufficiently concentrated that we may work in a flat sky
approximation and take $\theta$ to be a radial coordinate in the plane.
The radial component of the Fourier conjugate variable will be $\ell$.  The
azimuthally symmetric two-dimensional Fourier transform $B(\ell)$ of
$B(\theta)$ may be used to obtain the multipole coefficients according to
$B_\ell = B(\ell)$ with entirely negligible error.  We will thus work in
the flat sky approximation, but write $B_\ell$ (with $\ell$ in subscript)
to emphasize the correspondence to the multipole coefficients.

We consider contributions from the wing and from the core separately.  The
transform of the wing (which is truncated below $\theta_A$) is easily
obtained.  For the beam core, computation of the transform from the binned
radial profile requires us to interpolate the profile data $y_i$.  We
achieve this by expressing the real space beam as a sum of basis functions
which we know to have spatial frequency cut-offs near those determined by
the telescope optics in each frequency band.

The primary mirror has a diameter $D = 6$ m; this size limits the optical
response to spatial frequencies below $\ellmax{} \approx 2\pi D / \lambda$.
A natural basis for the Fourier space beam is then provided by the Zernike
polynomials $R_{2n}^0(\rho)$, where $\rho = \ell/\ellmax{}$.  (The Zernike
polynomials $R_n^m(\rho,\phi)$ are orthogonal and complete on the unit
disk; here we impose azimuthal symmetry and need only consider even $n$ and
$m=0$.)  We thus model the real space beam with basis functions
$f^n(\theta)$, for non-negative integer $n$, that are proportional to
inverse Fourier transforms of $R_{2n}^0(\rho)$:
\begin{align}
  f^n(\theta) & = \frac{J_{2n+1}(\theta \ellmax{})}{\theta\ellmax{}}.
\end{align}

In practice, for some value of $\ellmax{}$ and a mode count $n_{\rm mode}$
the beam model is expressed as
\begin{align}
  B(\theta) & = \sum_{n=0}^{n_{\rm mode}-1} a^n f^n(\theta),
\end{align}
where the coefficients $a^n$ are fit parameters.  The model is fitted to
the data by minimizing the $\chi^2$ of the residuals $y_i - B(\theta_i)$,
accounting for the full covariance of the $y_i$ measurements.  Values of
$\ellmax{}$ and $n_{\rm mode}$ are adjusted to obtain a fit with the
minimum number of modes necessary that gives a $\chi^2$ per degree of
freedom equal to unity.  The fit yields coefficients $a^n$ and the
covariance matrix of the errors, $\langle \delta a^n \delta a^{n'}
\rangle$.

Details of the fitting parameters, including $\ellmax{}$, $n_{\rm mode}$
and reduced $\chi^2$ for each season and array, are presented in
Table~\ref{tab:beam_props}.  The choice of basis functions results in a
satisfactory expansion of the beam using approximately half as many modes
as data points.

After fitting the coefficients $a^n$ in real space, we obtain the
corresponding beam transform,
\begin{align}
  B_\ell & = \sum_n a^n F^n_\ell + w_\ell,
\end{align}
where $F^n_\ell$ are the Fourier transforms of $f^n(\theta)$, truncated
above $\theta_A$, and $w_\ell$ is the Fourier transform of the extrapolated
wing, from $\theta_A$ to infinity.

\begin{figure*}[ht]
\begin{center}
\resizebox{\textwidth}{!}{\plotone{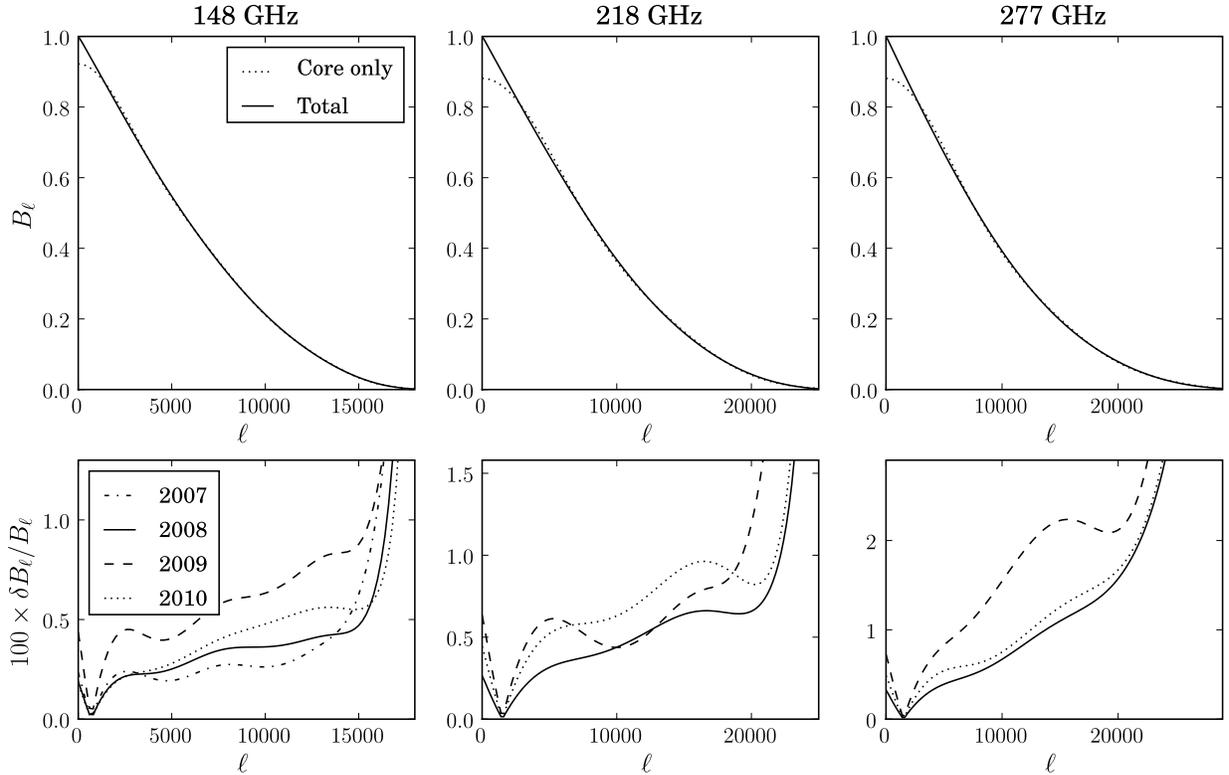}}
\caption{ \emph{Upper panels:} Beam transforms $B_\ell$ for each array.
  Beams shown are for the 2008 season.  The total beam is the sum of the
  contribution from the core of the beam, and the extrapolated wing.
  \emph{Lower panels:} Diagonal error from the covariance matrix for the
  renormalized beam.  The beam normalization has been fixed at calibration
  $\ell$ = 700 (1500) for the \arone{} (\artwo{} and \arthree{}) array(s),
  as described in \secref{sec:covariance}.  Note that these curves show
  uncertainty, not systematic trends, and they do not include additional
  uncertainty from the empirical corrections of
  \secref{sec:effective_beam}.  The full covariance matrix shows an
  anti-correlation between the beam error at angular scales above the
  calibration $\ell$ and beam error below the calibration $\ell$.  The
  window function is $B_\ell^2$ and thus its fractional error is $2\delta
  B_\ell/B_\ell$.  }
\label{fig:beam_panels}
\end{center}
\end{figure*}

The resulting beam transforms are shown in Figure~\ref{fig:beam_panels}.
While the beams for the 2008 season are shown, the beam features are not
substantially different between seasons.  The figure shows the separate
contributions of the core and wing to the beam transform.  The wing
contributes significantly only at low $\ell$ (i.e., below 1000 at \arone{},
and below 2000 at \artwoX{} and \arthree{}).

While this is very similar to the procedure described in
\cite{hincks/etal:2010}, in this work we include covariance between radial
profile data points, and between the radial profile and the wing fit
parameter.  This results in a natural propagation of beam errors on all
spatial scales into the fitted beam transform and its $\ell$-space
covariance matrix.  Our treatment of beam errors is discussed in the next
section.

\subsection{Beam and Calibration Covariance}
\label{sec:covariance}

\newcommand\expect[1]{\left\langle #1 \right\rangle}

Using the covariance matrix of the fit coefficients obtained in
\secref{sec:radial_profiles}, we also obtain the covariance in $\ell$
space:
\begin{align}
  \Sigma_{\ell\ell'} & \equiv 
  \expect{ \delta B_\ell \delta B_{\ell'} }\nonumber\\
  &=
  \sum_{n,n'} F^n_\ell F^{n'}_{\ell'} \expect { \delta a^n \delta a^{n'} }
  + \expect{ \delta w_\ell \delta w_{\ell'}}\nonumber\\
  & \quad + \sum_n F_\ell^n \expect{ \delta a^n \delta w_{\ell'} }
  + \sum_n \expect{ \delta w_\ell \delta a^n } F_{\ell'}^n.
\end{align}
This covariance includes contributions from the wing fit error and the
correlation between the wing fit error and the coefficients $a^n$.

This covariance matrix is an $\ell$-space representation of the covariant
features in the radial profiles that contribute to the $B_\ell$.  However,
because the calibration of the CMB survey maps is established at a
particular angular scale, we must recast the beam covariance into an
appropriate form.  This is effectively the same procedure applied in
\cite{page/etal:2003}.

The absolute calibration of the ACT maps at \arone{} is obtained by
cross-correlation of the 2008 Southern and 2010 Equatorial maps with the
WMAP 7-year maps \citep{jarosik/etal:2011} at 94~GHz of the same sky
region, over angular scales with $300 < \ell < 1100$
\citep{hajian/etal:2011, das/etal:2013}.  This leads to an absolute
calibration of the ACT maps centered near $\ell = 700$.  The 2009 season
Equatorial maps at \arone{} are then calibrated, over $500 < \ell < 2500$,
to the 2010 season Equatorial maps.  The subsequent calibration of the
\artwo{} maps to the \arone{} maps is performed in the signal-dominated
regime with $1000 < \ell < 3000$.  The \arthree{} maps have not yet been
calibrated to WMAP.  For each season and array, we factor out the beam
amplitude at effective calibration scale $\ell = L$, where $L = 700$ for
the \arone{} array, and $L = 1500$ for the \artwoX{} and \arthree{} arrays.

For a celestial temperature signal described by multipole moments
$T_{\ell,m}$, we measure a map $M_{\ell,m}$ in detector power units, which
is related to $T$ by
\begin{align}
  M_{\ell,m} & = G_\ell T_{\ell,m},
  \label{eqn:MfromT}
\end{align}
where $G_\ell \equiv G_0 B_\ell$ is the product of the beam (normalized
such that $B_{\ell=0} = 1$) and a global calibration factor $G_0$ that
converts CMB temperature to map units (i.e., to detector power units).

Our calibration to WMAP is a comparison of $M$ and $T$ at $\ell=L$ and is
thus a measurement of $G_L$ that is independent of our beam uncertainty.
This leads us to recast Equation~(\ref{eqn:MfromT}) as
\begin{align}
  M_{\ell,m} & = G_L b_\ell T_{\ell,m},
\end{align}
where
\begin{align}
  b_\ell \equiv \frac{B_\ell}{B_L}.
\end{align}
Since the errors in $b_\ell$ and $G_L$ are not correlated, the uncertainty
in $G_\ell$ is described by a covariance matrix $\Gamma_{\ell\ell'}$ that
is the sum of a calibration error from the measurement of $G_L$, and a term
due to correlated error in the $b_\ell$:
\begin{align}
  \Gamma_{\ell\ell'} & = b_\ell b_{\ell'}(\delta G_L)^2
  + \frac{G_L^2}{B_L^2} \left[
  \Sigma_{\ell\ell'} - b_\ell\Sigma_{L \ell'} \right.\nonumber\\
 & \left. \phantom{0000000000000000} - \Sigma_{\ell L}b_{\ell'} 
  + b_\ell b_{\ell'}\Sigma_{LL}
  \right].
\end{align}
The term in square brackets is the normalized beam covariance that
accompanies ACT data releases.  The diagonal beam error (i.e., the square
root of the diagonal entries of the normalized beam covariance) is shown
for each season and array in Figure~\ref{fig:beam_panels}.  At high $\ell$,
the error in the effective season beam is dominated by an empirical
map-based correction, which is not included here (see
\secref{sec:effective_beam}).

While the fitted beam and covariance are an accurate description of the
binned radial beam profile, they must be corrected for a number of
systematic effects prior to being used to interpret ACT maps.

\subsection{Correction for Systematics}
\label{sec:corrections}

The beams computed in the preceding section are corrected for a number of
systematic effects that would otherwise bias the resulting transforms
relative to the true telescope beam.  These are briefly described below.

\subsubsection{Mapping Transfer Function}
\label{sec:sys_mapping}

Because our map-making procedure includes time domain high-pass filtering,
we might expect poor reproduction of large spatial scales.  We study the
mapping transfer function by injecting a simulated signal into telescope
time stream data and comparing the output map to the input map in Fourier
space.  The transfer function deviates significantly from unity only on
angular scales larger than 20\arcmin{}.  The wing fit is performed at
somewhat smaller radii than this, and we have confirmed that the wing fit
and extrapolation reproduces the input signal, on all scales, to 0.1\%.

A very small ($\approx$0.1\%) correction due to the 3.5\arcsec{}
pixelization of the planet maps is applied by dividing the beam transform
by the azimuthal average of the analytic pixel window function.

\subsubsection{Radial Binning of Planet Maps}

The binning of the planet map pixel data into annuli has a slight impact on
the inferred beam transform.  This is quantified by evaluating the harmonic
transform of data points taken from a model of the radial beam profile, and
comparing it to the harmonic transform of points that include simulated
binning of the radial beam profile.  This resulting correction is
approximately 1\% at $\ell = 10000$.

\subsubsection{Saturn Disk and Ring Shape}

The Saturn angular diameter of approximately 8\arcsec{} is large enough
that we do not treat it as a point source.  For each season and array, we
deconvolve Saturn's shape assuming that it is a disk with solid angle equal
to the mean solid angle for all Saturn observations that contribute to the
mean instantaneous beam measurement.  This correction is approximately 2\%
at $\ell = 10000$.

While Saturn's rings complicate the spatial distribution of the planetary
signal, these effects are negligible at $\ell < 20000$, aside from an
overall change in average brightness, and need not be considered in the
deconvolution procedure.

\subsection{Mean Instantaneous vs. Effective Beam}
\label{sec:effective_beam}

After applying the corrections described in \secref{sec:corrections}, the
resulting beam describes the telescope response to a point-like radiation
source with approximately Rayleigh-Jeans spectrum, averaged over the focal
plane of each array.  We refer to this beam as the ``mean instantaneous
beam,'' to emphasize that the combination of observations taken at many
different times may lead to an effective window function that is different
from the instantaneous one.

For use in particular contexts, we compute an effective beam that includes
corrections for alternative frequency spectra, and for pointing variation
or other cumulative effects resulting from the combination of observations
taken at many different times.

We apply a simple first-order correction to obtain the effective beam for
different radiation spectra.  For radiation with a band effective frequency
$\nu$, the beam is taken to be
\begin{align}
  B'(\ell) & = B(\ell \nu_{\rm RJ} / \nu),
\end{align}
where $\nu_{RJ}$ is the effective frequency for radiation with a
Rayleigh-Jeans (RJ) spectrum.  Effective frequencies for various spectral
types are presented in \cite{swetz/etal:2011}.

In the absence of planetary sources, the absolute pointing registration of
individual ACT observations is not known.  Since each pixel of the ACT
survey maps contains contributions from data acquired on many different
nights throughout each season, the resulting season effective beam for
these maps is less sharp than the instantaneous beam obtained from
carefully aligned individual planet observations.

Pointing repeatability can be estimated from the variation in apparent
planet positions, relative to expectations, in each season and array.  As
summarized in Table~\ref{tab:pointing}, these indicate that the
repeatability of telescope pointing is at the $\sigma_{\rm planet} =
3\arcsec{}$ to 8\arcsec{} (RMS) level.  Pointing variation in CMB
observations may in principle be smaller than $\sigma_{\rm planet}$, since
CMB observations are performed at the same azimuth angles each night, while
planets observations span a wider range of azimuths.

In addition to pointing repeatability, the season effective beam may also
be diluted due to errors in the global pointing model, which is used to
combine observations made at different azimuth angles.  These global
adjustments are estimated using point source positions measured in
(non-cross-linked) maps made from only rising or only setting data.

While pointing variance and global alignment may contribute to dilution of
the season beam, we must also acknowledge that the beam may change slightly
over the course of the season in a way that is not captured by the Saturn
observations.  (Changes in the beam over the course of the season, and the
resulting impact on calibration, are assessed using Uranus observations in
\secref{sec:calibration}.)  We thus seek an empirical measure of the
difference between the season effective beam and the instantaneous beam
obtained from Saturn observations.

We parametrize the difference between the season effective beam
$B_\ell^{\rm eff}$ and the instantaneous beam $B_\ell$ as a Gaussian in
$\ell$:
\begin{align}
  B_\ell^{\rm eff} & = B_\ell \times e^{-\ell(\ell+1) V/2}.
\end{align}
If the correction is interpreted as arising purely from Gaussian pointing
error, then $V$ is simply the pointing variance in square radians.  While
this motivates the form of the correction, our broader interpretation
permits $V$ to differ from the value expected based on pointing variance
measured from planets, or to be less than zero if the season effective beam
is somewhat sharper than the ensemble of Saturn observations indicates.

We measure $V$ for each season and array using bright point sources in the
season CMB survey maps.  Point sources with signal to noise ratios greater
than 10 are identified using a matched filter.  A beam model is fit to the
map in the vicinity of each source, producing a value of $V$.  The weighted
mean and error of these individual fits, for each season and array, are
used to form the effective beam for use with the survey maps.  While this
point source analysis gives us a secondary probe of the beam core, we note
that the Saturn observations are still key for measuring the behavior in
the wings of the beam.

The values of the season correction parameter $V$ are presented in
Table~\ref{tab:pointing}.  The measurement of $V$ is somewhat susceptible
to changes in the fitting parameters and thus the quoted errors have been
inflated from the formal values by factors of 2 and 6 for the \arone{} and
\artwo{} arrays, respectively.  While values of $V$ for the \arone{} array
are very similar between seasons, values for \artwo{} are more variable.
The analysis of Uranus observations in \secref{sec:brightness_measurements}
shows that in 2010 the telescope beam was more sharply focused, relative to
the mean beam estimated from Saturn, during most of the observing season
(see \secref{sec:brightness_measurements}); this results in a negative
value for $V$ at \artwo{}.

\renewcommand\mytitle{ Pointing variance and beam correction parameters.  }
\renewcommand\mycaption{ Pointing variance $\sigma_{\rm planet}^2$ is
  measured from planet observations, and the season effective beam
  correction parameter $V$ is measured from point sources in the
  full-season CMB survey maps, where available.  The parameter $V$ is used
  to correct the high-$\ell$ beam for use with the survey maps; the
  resulting attenuation or inflation of the beam at $\ell = 5000$ is
  provided for reference.  The covariant error due to this correction is
  included in the season effective beam covariance matrix.  }
\begin{deluxetable}{c c c c c}
\tablecaption{\mytitle{}}
\tablehead{
        &       & $\sigma^2_{\rm planet}$ & $V$ & Attenuation     \\
 Season & Array &  (arcsec$^2$) &  (arcsec$^2$) & at $\ell=5000$ \\
}
\startdata
2007 & \arone{} & $$\phantom{-}$\phantom{0}53 \pm 12$ &        --        &        --       \\
\cline{1-5} \vspace{-.1cm} & & \\
2008 & \arone{} & $$\phantom{-}$\phantom{0}24 \pm \phantom{0}4$ & $$\phantom{-}$\phantom{0}25 \pm 10$ & $$\phantom{-}$0.993 \pm 0.003$\\
 & \artwo{} & $$\phantom{-}$\phantom{0}23 \pm \phantom{0}5$ & $$\phantom{-}$143 \pm 51$ & $$\phantom{-}$0.962 \pm 0.013$\\
 & \arthree{} & $$\phantom{-}$\phantom{0}11 \pm \phantom{0}2$ &        --        &        --       \\
\cline{1-5} \vspace{-.1cm} & & \\
2009 & \arone{} & $$\phantom{-}$\phantom{0}58 \pm \phantom{0}9$ & $$\phantom{-}$\phantom{0}26 \pm \phantom{0}6$ & $$\phantom{-}$0.993 \pm 0.002$\\
 & \artwo{} & $$\phantom{-}$\phantom{0}43 \pm \phantom{0}8$ & $$\phantom{-}$\phantom{0}46 \pm 27$ & $$\phantom{-}$0.988 \pm 0.007$\\
 & \arthree{} & $$\phantom{-}$\phantom{0}49 \pm 18$ &        --        &        --       \\
\cline{1-5} \vspace{-.1cm} & & \\
2010 & \arone{} & $$\phantom{-}$\phantom{0}31 \pm \phantom{0}3$ & $$\phantom{-}$\phantom{0}29 \pm \phantom{0}6$ & $$\phantom{-}$0.992 \pm 0.002$\\
 & \artwo{} & $$\phantom{-}$\phantom{0}18 \pm \phantom{0}2$ & $\phantom{00}$-$5 \pm 23$ & $$\phantom{-}$1.002 \pm 0.006$\\
 & \arthree{} & $$\phantom{-}$\phantom{0}25 \pm \phantom{0}3$ &        --        &        --       
\label{tab:pointing}
\tablecomments{\mycaption{}}
\enddata
\end{deluxetable}

\section{Planet Brightness Measurements}
\label{sec:calibration}

In this section we use observations of Uranus and Saturn to study the
variation in the system gain due to changes in water vapor level and
telescope focus.

\subsection{Planet Amplitudes}
\label{sec:brightness_measurements}

\newcommand{\true}{{(0)}}
\newcommand{\thetav}{\vec{\theta}}
\newcommand{\alphahat}{\widehat{\alpha}}
\newcommand{\betahat}{\widehat{\beta}}

We characterize the apparent brightness of a planet in a map by examining
the ratio of the Fourier transform of the map to expectations based on the
beam and planet shape.  While ultimately equivalent to procedures that
involve measuring source peak heights in filtered real-space maps, the
Fourier space approach emphasizes the role of spatial filtering in
providing precise planet amplitude measurements.  It also allows us to
quantify differences between the planet shape and expectations based on the
beam measurements in a physically meaningful way.

Once again working in the flat sky limit, we let $\thetav{}$ represent a
vector in the plane of the sky centered on the planet.  The Fourier
conjugate variable is denoted as $\ellv{}$ and is written in subscript.
The signal from a planet is modeled as a circular disk of uniform RJ
temperature $T^P$.  For each planet observation, we use the total disk
solid angle $\Omega^P$ to compute angular radius $\theta^P \equiv
(\Omega^P/\pi)^{1/2}$.  The Fourier transform of the planetary signal has
components
\begin{align}
  T_\ell & = T^P \Omega^P_\ell,
\end{align}
where $\Omega^P_\ell \equiv \Omega^P \times (2J_1(\ell \theta^P) / (\ell
\theta^P))$ is the Fourier transform of a disk of radius $\theta^P$.  For
Uranus, $\Omega^P_\ell \approx \Omega^P$ at the ACT angular scales; for
Saturn, the finite disk size cannot be neglected, but the effects due to
its oblateness are negligible.  Planetary radii are taken at the 1 bar
atmospheric surface, as reported in \citet{archinal/etal:2009}.  The Saturn
equatorial radius is $60268 \pm 4$ km and the polar radius is $54364 \pm
10$ km; the Uranus equatorial radius is $25559 \pm 4$ km and the polar
radius is $24973 \pm 20$ km.

Adapting equation~(\ref{eqn:MfromT}) to the flat sky, the ACT map
$M(\thetav{})$ of the planetary signal will have Fourier components
\begin{align}
  M_\ellv{} & = G_0 B_\ell T^P \Omega^P_\ell.
\end{align}

For each planet map, we compute the amplitude $M_\ell$ of the planet at
multiple bins in $\ell$.  The map is centered, apodized, and Fourier
transformed.  To suppress atmospheric contamination, which contributes
faint horizontal streaks to the map, we mask Fourier components having
$|\ell_x| < 1700$\footnote{No such masking or filtering is performed on the
  Saturn maps prior to their use in the determination of the telescope
  beam; the masking is more important for Uranus maps, because the signal
  is weaker.}.  The remaining complex Fourier amplitudes are averaged
inside annuli of width $\Delta \ell \approx 600$ and the imaginary part is
discarded.  We then compute the ratio of $M_\ell$ to $B_\ell \Omega^P_\ell$
(which has been corrected to account for the apodization and masking
applied to obtain $M_\ell$).

The result is a curve of planet amplitude measurements,
\begin{align}
  \alphahat{}_\ell & \equiv \frac{M_{\ell}}{B_\ell \Omega^P_\ell},
\end{align}
for $2000 < \ell < 14000$.  In the absence of any gain or beam variation,
$\alpha_\ell$ should be equal to $G_0 T^P$, the apparent brightness of the
planet in detector power units.

\begin{figure}[ht]
\begin{center}
\resizebox{3.5in}{!}{\plotone{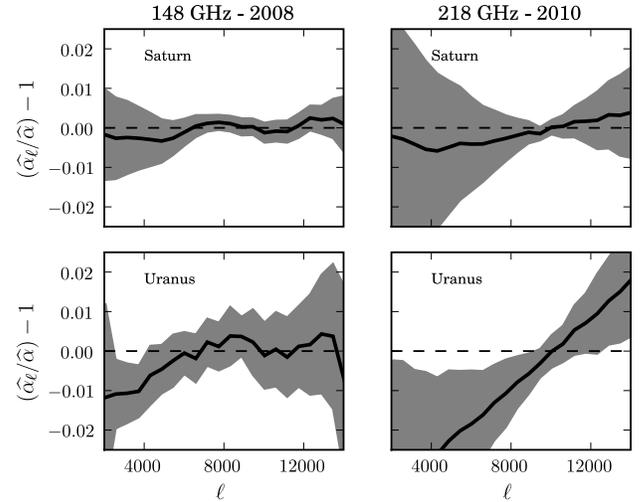}}
\caption{ Variation of planet amplitude measurements with angular scale.
  Each panel shows mean fractional deviation of $\alphahat{}_\ell$ relative
  to $\alphahat{}$ (solid line) and the standard deviation of all
  individual $\alphahat{}_\ell/\alphahat{}$ curves (grey band) for a
  particular season, array, and planet.  See
  \secref{sec:brightness_measurements} for definitions of
  $\alphahat{}_\ell$ and $\alphahat{}$.  We show results from \arone{} 2008
  (left panels), where there is a high degree of consistency between the
  planet observations and the season mean beam.  We also show results for
  \artwo{} 2010 (right panels), where there is somewhat more variance in
  the curves (which is typical of the \artwo{} array) and where the Uranus
  observations deviate significantly from the mean beam (which is seen in
  all arrays in 2010).  Differences between Uranus and Saturn are
  indicative of 2\% changes in the beam focus during the season; such
  changes have been accounted for in the season effective beam by fitting a
  correction parameter to maps of point sources in the season maps as
  described in \secref{sec:effective_beam}. }
\label{fig:alpha_fits}
\end{center}
\end{figure}

The mean and variance of $\alphahat{}_\ell$ curves for selected seasons and
arrays are shown in Figure~\ref{fig:alpha_fits}.  The $\alphahat{}_\ell$
curves are not in all cases consistent with a constant value.  This may be
attributed to variations in the focus from night to night.  To parametrize
these deviations and to make possible an assessment of the impact they
might have on the system calibration, we fit parameters $\alphahat{}$ and
$\betahat{}$ for each $\alphahat{}_\ell$ curve according to the following
model:
\begin{align}
  \alphahat{}_\ell & = \alphahat{} \left( 1 + \betahat{} 
  \frac{\ell - \bar{\ell}}{5000} \right).
  \label{eqn:alpha_ell}
\end{align}
The mean angular scale $\bar{\ell}$ is taken as the center of the fitting
range.  The parameters $\alphahat{}$ and $\betahat{}$ represent the mean
amplitude of the planet and the deviation from expected focus of the beam,
respectively.  For example, $\betahat{} = 0.01$ would indicate that the the
planetary signal is stronger by 1\% at $\ell \approx 10000$ relative to
$\ell \approx 5000$, given the mean season beam, and thus the telescope
beam at the time of observation was slightly more sharply focused.  We
would expect observations associated with negative $\betahat{}$ to also
have a smaller $\alphahat{}$ than expected, since a less sharply focused
beam will have poorer overall efficiency.

The extraction of amplitudes $\alphahat{}$ and focus parameters
$\betahat{}$ allows us to quantify deviations of focus over the season, and
to estimate the effects of such deviations on the overall system
calibration.  This is accomplished in the next section, where a model for
system calibration is obtained from the brightness and focus measurements
of Uranus.

\subsection{Calibration Parameters from Uranus Observations}
\label{sec:cal_params_uranus}

\newcommand\tauw{\tau_{\rm w}}
\newcommand\taud{\tau_{\rm d}}
\newcommand\taux{\tau_{\rm x}}

The measurements of $\alphahat{}$ and $\betahat{}$ from Uranus
observations, along with measurements of the precipitable water vapor (PWV)
level for most of these observations, permit us to characterize the impact
of PWV level and focus variation on the telescope calibration.  This
analysis does not require that the brightness temperature of Uranus be
precisely known, but assumes that the brightness is roughly constant over
the observing season.  (We do not perform the same analysis on Saturn,
because the effective brightness varies with the ring opening angle; see
\secref{sec:saturn_temp}).

The Atmospheric Transmission at Microwaves model
\citep[ATM;][]{pardo/etal:2001} provides estimates of atmospheric emission
and absorption at millimeter and sub-millimeter wavelengths.  The ACT
frequency bands avoid strong molecular resonances and are thus primarily
susceptible to continuum emission and absorption due to water vapor.  For
the purposes of calibration we are interested in the atmospheric
transmission in a given frequency band.  The transmission at the ACT
observation altitude $\theta_0$ is written as
\begin{align}
  T & = \exp \left(-(\tauw{} w + \taud{}) / \sin \theta_0 \right),
\end{align}
where $w$ is the precipitable water vapor column density at zenith and
$\tauw{} w$ and $\taud{}$ are the ``wet'' and ``dry'' optical depths at
zenith.  The parameters $\tauw{}$ and $\taud{}$ vary negligibly over the
range of temperatures and pressures experienced at the telescope site.

Because planet observations discussed in this work are obtained at the same
fiducial altitude used for CMB observations, the effects of $\taud{}$ are
common to both observations and cannot be measured from the planet
observations.  We define the overall system gain $G_0$ such that it applies
at the typical PWV level of $w_0 = 0.44$ mm.  The $w$-dependent gain factor
is then
\begin{align}
  G_0(w) & = G_0 \exp\left(-(\tauw{} + \taux{})(w - w_0) / \sin\theta\right).
\end{align} 
Here $\tauw{}$ is as provided by the ATM model, while $\taux{}$ is a fit
parameter that could indicate a deviation from the ATM model but is more
likely due to a bias in the detector calibration procedure that is
sensitive to sky loading.  Estimates of $\tauw{}$ for the ACT bands were
computed using the ATM model by \cite{marriage:2006}; these values are
provided for reference in \autoref{tab:pwv_fits}.

To assess the impact of small variations in telescope focus on planet
calibration, we include a dependence on the difference between the focus
parameter $\betahat{}$ (defined in Equation~(\ref{eqn:alpha_ell})) and the
season mean focus parameter $\bar{\beta}$.  Because a planet was observed
approximately once per night, we take $\bar{\beta}$ to be the average of
measured $\betahat{}$ for all planet observations for the array and season
under consideration.  We adopt an exponential form out of convenience;
calibration variation is small so a linear correction would be equivalent.

The full model for the Uranus brightness data $\alphahat{}$, $\betahat{}$
is then given by
\begin{align}
  \alpha & = T^U G_0 e^{m (\beta-\bar{\beta})} \times \nonumber\\
    & \quad\quad \exp\left(-(\tauw{}+ \taux{})(w - w_0) /
  \sin\theta\right),
  \label{eqn:alpha_model}
\end{align}
where the fit parameters $T^U G_0$, $m$, and $\taux{}$ represent overall
calibration, sensitivity to variations in telescope focus, and sensitivity
to PWV beyond predictions based on ATM, respectively.

This model is fitted to the Uranus brightness amplitude data for each
season and frequency array.  The parameter values are presented in
Table~\ref{tab:pwv_fits}.  The residuals have scatter of approximately 2\%,
2\% to 6\%, and 6\% in the \arone{}, \artwo{} and \arthree{} arrays
respectively; this scatter is not explained by uncertainties in planet
amplitude measurement.  Values obtained for $\taux{}$ are in good agreement
between seasons for the \arone{} and \artwo{} arrays.  They are
inconsistent with zero (at the $\approx 2$--$\sigma$ to 4--$\sigma$ level),
which we attribute to systematic detector calibration error.  For the
\arthree{} array there is some slight disagreement between the 2009 and
2010 results, which we attribute to the low number of observations in 2009.
The values of $\taux{}$ are used in combination with $\tauw{}$ to correct
the time-ordered data for opacity prior to creation of full-season survey
maps, as described in \cite{dunner/etal:2013}.

While for some seasons and arrays we obtain better fits to the data by
including the parameter $m$, the impact of this additional parameter on
inferred overall system calibration is small.  To show this, we refit the
model of equation (\ref{eqn:alpha_model}) with parameter $m$ fixed to 0.
The resulting calibration factor $T^U G_0^{(m=0)}$ may be compared to the
case where $m$ was free to vary.  The differences are presented in
Table~\ref{tab:pwv_fits}; the change in overall calibration is 0.6\% or
smaller, and always less than the 1-$\sigma$ error on $T^U G_0$, with the
exception of the 2009 data for the \artwo{} array.  This exceptional case
also has a large (9\%) uncertainty in its absolute WMAP-based calibration.

The fit values of $T^U G_0$ may be used to provide an absolute calibration
of each array (should $T^U$ be known) or to provide a measurement of the
Uranus disk temperature based on an independent calibration of $G_0$.  The
latter possibility is discussed in the next section, leading to the
measurements of Uranus and Saturn brightness temperatures presented in
\secref{sec:brightness}.

\renewcommand\mytitle{ Properties of planetary data and calibration model
  fit parameters by season and array.  }
\renewcommand\mycaption{ The opacity parameters $\tauw{}$ from the ATM
   model are provided for reference.  Fit parameters $\taux{}$ and $m$
   parametrize sensitivity of Uranus apparent brightness to PWV and focus
   parameter, respectively.  The mean and standard deviation of the focus
   parameter $\beta$ for all planet observations in each season is provided
   for reference.  The overall calibration shift $G_0^{(m=0)}/G_0 - 1$ is
   the systematic calibration change if $m$ is fixed to zero instead of
   being free to vary.  } \begin{deluxetable}{l c c c}
\tablecaption{\mytitle{}}
\tablehead{
 & \multicolumn{3}{c}{Array}\\
 Quantity& \arone{} & \artwo{} & \arthree{} 
}
\startdata
$\tau_{\rm w}$ (mm$^{-1}$)  & $0.019$ & $0.044$ & $0.075$\\
\cline{1-4} \vspace{-.1cm} & & \\
$\tau_{\rm x}$ (mm$^{-1}$) & & & \\
\quad 2007  & $$\phantom{-}$0.016 \pm 0.010$ &        --        &        --       \\
\quad 2008  & $$\phantom{-}$0.014 \pm 0.004$ & $$\phantom{-}$0.040 \pm 0.008$ & $$\phantom{-}$0.033 \pm 0.063$\\
\quad 2009  & $$\phantom{-}$0.009 \pm 0.005$ & $$\phantom{-}$0.021 \pm 0.017$ & $$-$0.008 \pm 0.041$\\
\quad 2010  & $$\phantom{-}$0.012 \pm 0.003$ & $$\phantom{-}$0.045 \pm 0.008$ & $$\phantom{-}$0.083 \pm 0.016$\\
\cline{1-4} \vspace{-.1cm} & & \\
\multicolumn{2}{l}{Error in $T^U G_0$ (\%)} & & \\
\quad 2007  & $\phantom{-}$0.9 &        --        &        --       \\
\quad 2008  & $\phantom{-}$0.2 & $\phantom{-}$0.4 & $\phantom{-}$1.3\\
\quad 2009  & $\phantom{-}$0.3 & $\phantom{-}$1.1 & $\phantom{-}$2.2\\
\quad 2010  & $\phantom{-}$0.2 & $\phantom{-}$0.4 & $\phantom{-}$0.8\\
\cline{1-4} \vspace{-.1cm} & & \\
$m$ & & & \\
\quad 2007  & $$\phantom{-}$0.7 \pm 0.6$ &        --        &        --       \\
\quad 2008  & $$\phantom{-}$0.5 \pm 0.3$ & $$-$0.5 \pm 0.6$ & $$\phantom{-}$1.2 \pm 1.3$\\
\quad 2009  & $$\phantom{-}$0.5 \pm 0.3$ & $$\phantom{-}$3.4 \pm 0.5$ & $$\phantom{-}$1.2 \pm 0.7$\\
\quad 2010  & $$\phantom{-}$0.8 \pm 0.1$ & $$\phantom{-}$0.5 \pm 0.2$ & $$\phantom{-}$1.5 \pm 0.4$\\
\cline{1-4} \vspace{-.1cm} & & \\
\multicolumn{2}{l}{$\beta$ mean (rms)} & & \\
\quad 2007  & $-$0.017 (0.010) &        --        &        --       \\
\quad 2008  & $\phantom{-}$0.002 (0.006) & $\phantom{-}$0.011 (0.010) & $\phantom{-}$0.011 (0.015)\\
\quad 2009  & $\phantom{-}$0.004 (0.011) & $\phantom{-}$0.008 (0.016) & $\phantom{-}$0.007 (0.020)\\
\quad 2010  & $\phantom{-}$0.015 (0.015) & $\phantom{-}$0.021 (0.019) & $\phantom{-}$0.017 (0.019)\\
\cline{1-4} \vspace{-.1cm} & & \\
\multicolumn{2}{l}{$\left(G_0^{(m=0)}/G_0 - 1\right)$ (\%)} & & \\
\quad 2007  & $-$0.6 &        --        &        --       \\
\quad 2008  & $\phantom{-}$0.1 & $-$0.3 & $\phantom{-}$0.5\\
\quad 2009  & $\phantom{-}$0.2 & $\phantom{-}$1.4 & $\phantom{-}$0.4\\
\quad 2010  & $\phantom{-}$0.3 & $\phantom{-}$0.2 & $\phantom{-}$0.4
\enddata
\label{tab:pwv_fits}
\tablecomments{\mycaption{}}
\end{deluxetable}

\subsection{Relation to WMAP-Based Calibration}

The absolute calibration of the ACT full-season CMB survey maps is obtained
through cross-correlation in $\ell$-space \citep{hajian/etal:2011} to the
94~GHz maps from WMAP7 \citep{jarosik/etal:2011}.  As described by
\cite{das/etal:2013}, the WMAP maps are used to calibrate the ACT 2008
Southern and 2010 Equatorial field maps at \arone{} directly; the latter
maps are then used to calibrate the \arone{} map for 2009 and the \artwo{}
maps for all seasons.

While the data entering the survey maps are corrected for atmospheric
opacity, they have not been corrected for any kind of time variation in the
telescope beam.  Instead, we have obtained an effective season beam
suitable for use with the survey maps by comparing our mean instantaneous
beam to point sources in the survey maps, as described in
\secref{sec:effective_beam}.  This correction has negligible effect at low
$\ell$ where the calibration to WMAP takes place.  Furthermore, the WMAP
calibration should be associated with the telescope gain at the
season-average focus parameter.  Thus the calibration factor obtained from
the WMAP calibration is compatible with $G_0$ as defined by
equation~(\ref{eqn:alpha_model}).

\section{Brightness Temperatures of Uranus and Saturn}
\label{sec:brightness}

Using the WMAP-based calibration of the ACT survey maps at \arone{} and
\artwo{}, we convert the Saturn and Uranus amplitude measurements to
calibrated brightness temperatures.  A geometrically detailed,
two-parameter model for Saturn disk and ring temperatures is fit to the
Saturn brightness data.  Because CMB survey maps for the \arthree{} array
have not yet been calibrated to WMAP, we do not consider that array in this
analysis.

In what follows, we refer to both the Rayleigh-Jeans (RJ) and brightness
temperatures of the planet.\footnote{Our planet amplitude measurements
  include a deconvolution of the telescope beam and thus we do not work
  with the planet's ``antenna temperature'' directly.}  The RJ temperature
$T_{\rm RJ}$ is related to the brightness temperature $T_{\rm B}$ by
equating the specific intensity of an RJ spectrum and a blackbody spectrum
at the band effective frequency $\nu_{\rm eff}$:
\begin{align}
  B_{\nu_{\rm eff}}(T_{\rm B}) & = 2 \nu_{\rm eff}^2 k_{\rm B} T_{\rm RJ} / c^2,
\end{align}
where $B_\nu(T)$ is the blackbody spectral radiance.  The spectra of Saturn
and Uranus are each sufficiently close to the RJ limit that we take the
source effective frequencies to be the RJ band centers of 149.0~GHz and
219.1~GHz \citep{swetz/etal:2011}.

\subsection{Uranus}
\label{sec:uranus_temp}

Based on our fitted values of $T^U G_0$ (\secref{sec:cal_params_uranus})
and the absolute calibration of $G_0$ using WMAP data, we obtain
measurements of $T^U$.  Because the calibration factor is obtained from the
primary anisotropies of the CMB, we must account for the different
frequency spectrum of the planets relative to the CMB blackbody.  The
effective spectral index of the flux of the CMB in the ACT bands is 1.0
(0.0) in the \arone{} (\artwo{}) bands, while the spectral index of Uranus
is approximately 1.65 in both bands \citep[based on,
  e.g.,][]{griffin/orton:1993}.  The 3.5~GHz uncertainty in the ACT band
centers and the effective frequencies provided in \cite{swetz/etal:2011}
result in an additional 1.6\% (2.7\%) error in the Uranus brightness
determinations at \arone{} (\artwo{}).

The Rayleigh-Jeans temperatures obtained for each season and array are
presented in Table~\ref{tab:T_planet}.  The value quoted for the \arone{}
(\artwo{}) array is valid at effective frequency of 149.0 (219.1)~GHz.  All
RJ temperatures have been inflated by 0.5 (0.3) K to compensate for the
background provided by the CMB in our bands.  Brightness temperatures may
be computed from the RJ values by adding 3.5 (5.1) K.

The results are consistent between seasons (even after removing the
correlated error due to the frequency spectrum correction), with
uncertainty dominated by the overall instrument calibration uncertainty
rather than statistical uncertainty from planet brightness measurements.

From the RJ temperature measurements in each season, we compute a weighted
mean disk temperature for Uranus.  The WMAP-based calibration error is
correlated between the 2009 and 2010 seasons, because the 2009 map is
calibrated to the 2010 ACT map, and the ACT band center uncertainty is
correlated across all seasons.  We account for this in the mean and error
presented.  At \arone{}, we obtain a RJ (brightness) temperature of
\TUoneRJ{} (\TUoneB{}), with \TUoneE{} error.  At \artwo{}, we obtain
temperature \TUtwoRJ{} (\TUtwoB{}), with \TUtwoE{} error.

\newcommand\GO{G\&O}

We compare our Uranus temperature measurements to the widely used model of
\citet[][hereafter \GO{}]{griffin/orton:1993}.  The \GO{} model provides
brightness temperatures for Uranus and Neptune in the $\approx 100$ to
1000~GHz range based on precise measurements of brightness ratios of each
planet to Mars.  The absolute calibration is obtained by interpolating, in
the logarithm of the frequency, between the \cite{wright:1976} model for
Mars temperature at $3.5 \micron$ (857~GHz), and the \citep{ulich:1981}
model for Mars temperature at 90 GHz.  They adopt a 5\% systematic error
for this Ulich-Wright hybrid model.

The \GO{} model predicts a brightness temperature of 112 K at \arone{},
approximately 5\% larger than our measured value.  This suggests that the
Ulich-Wright model for Mars brightness is 5\% high at \arone{}.
\cite{weiland/etal:2011} have observed a similar 5\% discrepancy between
Mars temperatures and the Ulich model at 94 GHz.  At \artwo{}, the \GO{}
model gives a brightness temperature of 99 K, which is 1\% smaller than our
measured value; this difference is smaller than our calibration
uncertainty.

Long term studies of Uranus show a variation in mean disk brightness at
8.6~GHz \citep{klein/hofstadter:2006} and 90~GHz
\citep{kramer/moreno/greve:2008} on decade time scales. If this variation
is attributed to differences in the Uranus surface brightness at different
planet latitudes, it implies that the mean disk temperature is roughly 10\%
larger when the South pole, rather than the equator, is directed to towards
the observer.\footnote{The center of the projected planet disk, as observed
  from Earth, is called the sub-Earth point.  The Uranian latitude of this
  point is called the sub-Earth latitude.}  The ACT measurements over the
2008--2010 period span Uranus sub-Earth latitudes from 1\degree{} to
13\degree{}.  In comparison, the observations presented by \GO{} were made
in 1990--1992 and span sub-Earth latitudes from -70\degree{} to
-60\degree{}.  It is then possible that, rather than a difference in
calibration, Uranus was brighter in these bands at the time the \GO{} data
were taken.  However, if one accepts the 5\% downward recalibration of the
Ulich Mars model at 94 GHz, then one should accept a downward recalibration
(of 4--5\%) of the \GO{} data around \arone{}.  This brings the ACT and
\GO{} measurements into agreement, with no evidence for any variation in
Uranus brightness temperature due to sub-Earth latitude.

Our results at \arone{} are consistent with the analysis of
\cite{sayers/czakon/golwala:2012}, who use WMAP measurements of Uranus and
Neptune at 94~GHz \citep{weiland/etal:2011} to recalibrate the \GO{} model
at 143~GHz.  It is important to note that while our measurements rely on a
detailed understanding of the telescope beam, our absolute calibration
standard is the CMB, for which the brightness as a function of frequency is
known.  Our errors thus do not include any significant systematic
contribution due to interpolation of unknown source spectra into our
frequency bands.

\begin{figure}[ht]
\begin{center}
\resizebox{3.5in}{!}{\plotone{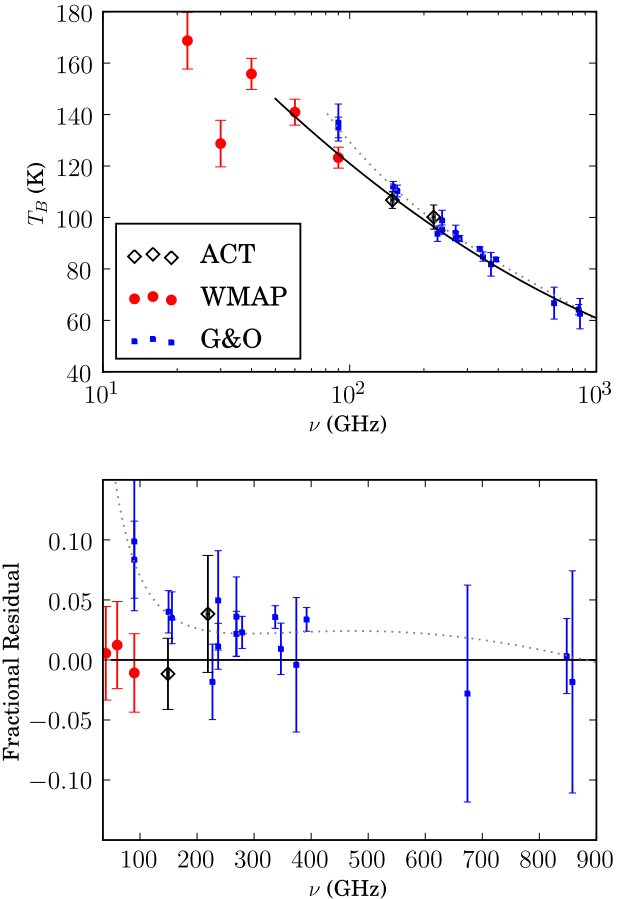}}
\caption{ Measurements of the Uranus brightness temperature from this work,
  \citet[WMAP]{weiland/etal:2011}, and \citet[G\&O]{griffin/orton:1993}.
  The \GO{} points include data from that work as well as
  \citet{ulich:1981} and \citet{orton/etal:1986}.  The dotted line shows
  the model of \GO{}.  \GO{} data are calibrated to a Ulich-Wright hybrid
  model for Mars brightness that interpolates between 90 and 857~GHz and
  carries an estimated 5\% systematic error.  The solid line is our
  best-fit empirical model, using only the two highest frequency points
  from WMAP, the two ACT points, and the \GO{} points above 600~GHz.}
\label{fig:uranus_summary}
\end{center}
\end{figure}

In Figure~\ref{fig:uranus_summary} we show our brightness temperature
measurements along with WMAP measurements below 100 GHz from
\citet{weiland/etal:2011} and with data used by \GO{} to fit their model in
the 90 to 1000 GHz range.

As discussed above, systematic error in the \GO{} model is dominated by
uncertainty in the Mars spectrum below 1000 GHz.  The ACT and WMAP data
provide new, absolutely calibrated temperature measurements between 20 and
220 GHz.  We thus fit a simple empirical model for the temperature using
the ACT, WMAP, and the higher frequency \GO{} data.  To avoid the strong
spectral feature at 30~GHz, we include only the two highest frequency bands
of WMAP.  We use the three data points in \GO{} above 600 GHz.  We model
the Uranus brightness temperature as a function of $\phi \equiv (\nu /
100~\rm{GHz})$ by
\begin{align}
  \frac{T^U}{1~\rm{K}} & = a_0 + a_1 \log_{10} \phi + a_2 \log_{10}^2 \phi,
\end{align}
and obtain best fit coefficients $(a_0, a_1, a_2) = (121, -78.2, 18.2)$.

The ACT, WMAP, and \GO{} brightness temperature measurements are shown in
Figure~\ref{fig:uranus_summary}, along with the \GO{} model and our
empirical model.

\renewcommand\mytitle{ RJ temperature measurements of Uranus and Saturn.  }
\renewcommand\mycaption{ Values for Uranus (\secref{sec:uranus_temp}) are
  shown for each season, as well as a weighted mean that takes full account
  of calibration covariance between seasons.  Saturn results
  (\secref{sec:saturn_temp}) are provided for both a two-component
  disk+ring model and a disk-only model.  All temperatures are RJ and have
  been corrected for the CMB to indicate the brightness that would be seen
  relative to an empty (T=0) sky.  To obtain thermodynamic brightness
  temperatures at \arone{} (\artwo{}), add 3.5 K (5.1 K) to the RJ values.
} \begin{deluxetable}{l c c c c c}
\tablecaption{\mytitle{}}
\tablehead{
 & \multicolumn{2}{c}{$T^{\rm RJ}$ (K)}\\
 & \arone{} & \artwo{} 
}
\startdata
Uranus (\S \ref{sec:uranus_temp}) & &\\
\quad 2008  & $101.5 \pm 2.0$ & $\phantom{0}93.2 \pm 2.3$\\
\quad 2009  & $103.8 \pm 2.2$ & $\phantom{0}91.5 \pm 8.6$\\
\quad 2010  & $105.7 \pm 2.1$ & $\phantom{0}97.6 \pm 2.6$\vspace{.2cm}\\
\quad Combined  & $103.2 \pm 2.2$ & $\phantom{0}95.0 \pm 3.1$\\
\cline{1-3} \vspace{-.1cm} & & \\
Saturn, W11 model (\S \ref{sec:saturn_temp}) & &\\
\quad $T_{\rm disk}$  & $133.8 \pm 3.2$ & $132.2 \pm 4.7$\\
\quad $T_{\rm ring}$  & $\phantom{0}17.7 \pm 2.2$ & $\phantom{0}12.3 \pm 4.0$\\
\quad [$\chi^2$ / d.o.f.]  & [7.6 / 5] & [6.5 / 5]\\
\cline{1-3} \vspace{-.1cm} & & \\
Saturn, single T model (\S \ref{sec:saturn_temp}) & &\\
\quad $T_{\rm disk}$  & $131.2 \pm 3.1$ & $127.5 \pm 4.4$\\
\quad [$\chi^2$ / d.o.f.]  & [12.0 / 6] & [11.2 / 6]
\enddata
\label{tab:T_planet}
\tablecomments{\mycaption{}}
\end{deluxetable}

\subsection{Saturn}
\label{sec:saturn_temp}

\newcommand\Tdisk{T_{\rm disk}}
\newcommand\Tring{T_{\rm ring}}

Measurements of the Saturn disk temperature are complicated by the presence
of the rings, whose dust both obscures the main disk and radiates at a
lower temperature.  The ACT observations of Saturn span ring opening angles
from -2 to 12 degrees and provide an opportunity to explore the separate
contribution of these two components to the total effective brightness of
the planet.

We follow closely the approach of \citet[hereafter W11]{weiland/etal:2011},
who apply a two-parameter model in which the opacity and geometry of the
rings are fixed, while the disk temperature and an effective temperature
for the rings are free to vary.  The ring dimensions and normal opacities
are taken from \cite{dunn/molnar/fix:2002}.  For each of the \arone{} and
\artwo{} arrays, we fit $\Tdisk{}$ and $\Tring{}$, which describe the mean
disk temperature and mean effective brightness temperature of the rings
(see W11 for the detailed definition), respectively.

\begin{figure}[ht]
\begin{center}
\resizebox{3.5in}{!}{\plotone{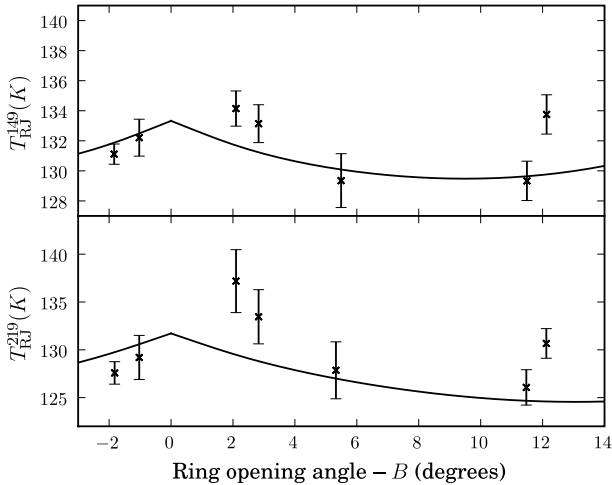}}
\caption{Binned data and resulting best-fit two-component model of
  effective Saturn brightness vs. ring opening angle ($B$) relative to
  observer.  The model is symmetric about $B=0$ by construction.  As the
  absolute value of $B$ increases, the increased radiation from the rings
  and decreased flux from the obscured disk leads to a local minimum in the
  total effective brightness at $B \approx $ 9\degree{} (13\degree{}) at
  149 GHz (219 GHz).  The model is fitted independently for each frequency
  band.  Error bars correspond to the error in the mean of observations
  contributing to each point, but fits include additional error due to
  calibration, which is covariant within each season and in some cases
  between seasons.  Approximate mean dates of observation are, from left to
  right: Nov/08, Dec/08, Jun/10, Apr/10, Dec/09, Dec/10, Dec/10.  The sign
  convention is such that $B<0$ corresponds to negative values of the
  sub-Earth latitude.  }
\label{fig:saturn_model}
\end{center}
\end{figure}

The measurements of $\alphahat{}$ and $\betahat{}$ for each Saturn
observation are converted to RJ temperatures, including corrections for PWV
level and focus parameter, and with the CMB-based calibration applied.
Because the scatter in the temperatures is not fully accounted for by the
errors in the amplitude measurements and correction parameters, we combine
observations made in single 15 to 30 day periods, taking the uncertainty in
each combined measurement to be the error in the mean of the contributing
data.  As a result, a small number of temporally isolated observations are
discarded.  The binned data points, and the best fit model, are shown in
Figure~\ref{fig:saturn_model}.

The fit of the W11 model includes a full accounting of the non-trivial
correlations in the calibration uncertainty of the data points.  The
resulting models are good fits, with $\chi^2$ per degree of freedom of
7.6/5 (6.5/5) in the \arone{} (\artwo) array.  The model fit parameters and
uncertainties are presented in Table~\ref{tab:T_planet}.

For comparison, we also fit a single temperature model to the binned Saturn
data, finding poorer fits and mean temperatures that are 1--$\sigma$ less
than the disk temperatures obtained in the full model.  These temperatures
and $\chi^2$ statistics are also presented in Table~\ref{tab:T_planet}.

\begin{figure}[ht]
\begin{center}
\resizebox{3.5in}{!}{\plotone{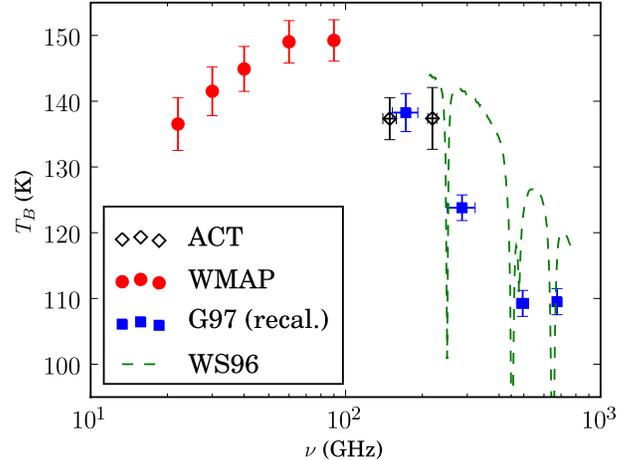}}
\caption{ Measurements of the Saturn disk brightness temperature from ACT,
  \citet[WMAP]{weiland/etal:2011}, and \citet[G97]{goldin/etal:1997}.
  Points from G97 have been recalibrated as described in the text.
  Frequency error bars on ACT and G97 points indicate bandwidth.
  Temperature error bars on G97 points do not include calibration
  uncertainty.  The green dashed line is a spectral model from
  \citet[WS96]{weisstein/serabyn:1996}, which shows proximity of absorption
  features to G97 points. }
\label{fig:saturn_summary}
\end{center}
\end{figure}

In Figure~\ref{fig:saturn_summary} we show the ACT disk temperature
measurements along with high precision results from W11 below 100~GHz and
from \citet[hereafter G97]{goldin/etal:1997} between 170 and 700~GHz.  The
G97 measurements have statistical errors at the 2\% level, but are
calibrated with reference to the same Mars model used by \GO{}.  We have
applied a rough recalibration factor to the G97 measurements using the WMAP
result \citep{weiland/etal:2011} that the \citet{ulich:1981} model for Mars
temperature is 5\% high at 94~GHz.  Our recalibration factor varies
linearly with the log of frequency from 0.95 at 90~GHz to 1 at 857~GHz.
With this calibration factor applied, the G97 measurement at 170 GHz and
ACT measurements at 149 and 219~GHz are consistent with a flat spectrum
over this frequency range.  It is difficult to comment further on the
continuum spectrum of Saturn based on the ACT and G97 data, due to the
proximity of the high frequency G97 bands to PH$_3$ resonances
\citep[e.g.,][]{weisstein/serabyn:1996}.

\section{Conclusion}

We have described the reduction of ACT planet observations for the purposes
of measuring and monitoring the telescope beam and gain parameters.  The
core of the beam has been well characterized by Saturn observations, and
studies of point sources in the map have provided an empirical correction
to account for pointing variance, alignment error, and slight changes in
beam focus.  Observations of Uranus have been used to measure and account
for changes in system gain due to atmospheric water vapor and focus
variation.

Based on hundreds of brightness measurements of Uranus, we have obtained
precise measurements of the disk brightness temperature at \arone{} and
\artwo{}.  In contrast to previous observations, our absolute calibration
standard is the CMB blackbody.  We have also obtained precise measurements
of the Saturn disk temperature, in the context of a simple two-component
model.\\


This work was supported by the U.S. National Science Foundation through
awards AST-0408698 and AST-0965625 for the ACT project, as well as awards
PHY-0855887 and PHY-1214379. Funding was also provided by Princeton
University, the University of Pennsylvania, and a Canada Foundation for
Innovation (CFI) award to UBC. ACT operates in the Parque Astron\'omico
Atacama in northern Chile under the auspices of the Comisi\'on Nacional de
Investigaci\'on Cient\'ifica y Tecnol\'ogica de Chile
(CONICYT). Computations were performed on the GPC supercomputer at the
SciNet HPC Consortium. SciNet is funded by the CFI under the auspices of
Compute Canada, the Government of Ontario, the Ontario Research Fund --
Research Excellence; and the University of Toronto.


\bibliographystyle{apj}

\end{document}